\documentclass[11pt]{article}
\usepackage{graphicx,color}
\usepackage{rotating}
\usepackage{epsfig,graphics,rotate,color}
\usepackage{wrapfig}
\usepackage{url}
\advance \topmargin by -\headheight
\advance \topmargin by -\headsep
\evensidemargin \oddsidemargin
\usepackage[T1]{fontenc}
\usepackage{times}

\begin{document} 
\begin{center}
\Large{Applications of Nanoparticles for Particle Physics: \\  A Whitepaper for Snowmass 2013}\\[0.25cm]
\large{Lindley Winslow,  \today}\\[0.25cm]
\end{center}

The last decade has been the decade of nanotechnology, a length scale which is of particular interest since it is here that we see the transition from the classical to the quantum world. In this transition to the quantum regime new phenomena appear that have proven valuable in a wide range of applications. This whitepaper focusses on the simplest nanotechnology, the spherical nanoparticles.  These are particles with diameters between 1-100~nm and they fall into two categories, metal nanoparticles and semiconductor nanoparticles. The latter are more often referred to as quantum dots.

It is the semiconducting nanoparticles (or quantum dots) whose properties have begun to be explored for applications to particle physics. In this size regime, the quantum dot stops acting like a bulk semiconductor and starts acting like an individual atom. This leads to interesting optical properties. Quantum dots absorb all light shorter than a characteristic wavelength and reemit this light in a narrow $\sim$25~nm resonance around this wavelength. This is a quantum confinement effect, and therefore the characteristic wavelength is directly proportional to the size of the quantum dot; larger dots have a longer characteristic wavelength. The chemical synthesis of the quantum dots allows for very good control of the size and therefore precise control of the emission wavelength of the solution.

The most common quantum dot is a core CdSe crystal, which has a size of 2~nm to 8~nm leading to emission spectra between 450~nm and 650~nm.   A shell of another crystal like ZnS is sometimes used to passivate secondary surface states which affect the optical properties. These are referred to as core-shell quantum dots. To obtain longer or shorter wavelength emission, other crystals are needed. For shorter wavelengths, CdS  can reach emission wavelengths of 380~nm. For longer wavelengths, CdTe reaches emission wavelengths out to 800~nm, and InAs pushes out into the infrared with maximum wavelengths of 1400~nm for InAs/CdSe core-shell quantum dots\cite{C0AN00233J}. Due to the toxicity of Cd, alternative compositions, such as ZnS and GdSe quantum dots, are actively being researched.

The chemical synthesis of quantum dots results in a colloidal solution. A coating of organic ligands is used to suspend the quantum dots in the solvent. The ligands are most often tuned for organic solvents, like toluene, but polar solvents, like water are also possible. Further chemistry can be done to embed the quantum dots into a solid material, thin film, gel, or hydrogel. They can also be transferred to a substrate for the construction of nanodevices. In colloidal solution or embedded in solids, gels, or thin films, the quantum dots act as perfectly tunable wavelength shifters, and therefore have obvious application to scintillator-based detectors. The response of quantum-dot-doped scintillator has been studied for gamma rays\cite{letant}, electrons\cite{campbell2006}, and thermal neutrons \cite{Wang2010186}.  Quantum-dot-doped scintillator is particularly interesting for neutrino physics\cite{qdot}. Beyond the wavelength-shifting properties, the crystal composition has interesting nuclear properties: Cd and Gd have high thermal neutron cross sections, making them good for antineutrino detection; while Cd, Se, and Te are candidate isotopes for neutrinoless double-beta decay searches.

There has been much work on being able to connect electrodes to quantum dots to form nanodevices. One of the first devices was a single-electron transistor made from a CdSe quantum dot\cite{qdot_trans}. A device that may prove very interesting to particle physics is the quantum dot photodetectors\cite{qdot_photodetect}. These nano-scale instruments are currently slower than photomultiplier tubes, with response times on the order of 200~ns, however like quantum dots as wavelength shifters, they offer unique flexibility in the tuning of the spectral response.  This is just one exciting device; the unique electric and optical properties of quantum-dot-based nanodevices may have many other interesting applications to particle physics.

Gold nanoparticles in colloidal suspension are the most common metallic nanoparticles and, like quantum dots, they have intriguing optical properties. The phenomenon here is a surface plasmon effect. For particles around 30~nm, light around 450~nm is absorbed while longer wavelengths are reflected giving the solution a red color. As the particles get larger, the absorption moves to higher wavelengths, and the reflected spectrum becomes more blue until finally the solution becomes clear as all visible light is reflected.  In addition to optical properties, gold nanoparticles are useful for making electrical connections in nanoscale devices.

The last class of nanoparticles are the magnetic nanoparticles. Iron oxide is common with variations including cobalt and nickel and shells of non-magnetic materials to passivate surface states. These nanoparticles are interesting because in the 5-50~nm range they exhibit superparamagnetism. The nanoparticles in this state act like paramagnets but with an even larger magnetic susceptibility. Like all nanoparticles, the main uses to date are medical applications. There is also some work being done developing them for use in data storage disks, which of course, would benefit particle physics. The novel chemistry and magnetic properties may also make for interesting applications in particle physics. 

Nanoparticles are being utilized in a wide variety of applications so it is logical that there would be applications to particle physics. The precision wavelength shifting that quantum dots provide has obvious application to particle detectors and is already being explored. Quantum-dot-based sensors show promise for future detector technology. Finally, the properties of gold and magnetic nanoparticles are fascinating, and though not as straightforward to apply, some more thought may lead to interesting applications in particle physics.  In general nanotechnology is advancing rapidly, and harnessing some of this momentum may lead to great advances in the next generation of particle physics experiments.

\bibliography{NSF_Winslow2012} 

\providecommand{\href}[2]{#2}\begingroup\raggedright\begin{thebibliography}{1}

\bibitem{C0AN00233J}
Q.~Ma and X.~Su, ``Near-infrared quantum dots: synthesis{,} functionalization
  and analytical applications,''
  \href{http://dx.doi.org/10.1039/C0AN00233J}{{\em Analyst} {\bfseries 135}
  (2010) 1867--1877}. \url{http://dx.doi.org/10.1039/C0AN00233J}.

\bibitem{letant}
S.~E. LŽtant and T.-F. Wang, ``Semiconductor quantum dot scintillation under
  ?-ray irradiation,'' \href{http://dx.doi.org/10.1021/nl0620942}{{\em Nano
  Letters} {\bfseries 6} no.~12, (2006) 2877--2880}.
  \url{http://pubs.acs.org/doi/abs/10.1021/nl0620942}. PMID: 17163723.

\bibitem{campbell2006}
I.~Campbell and B.~Crone, ``Quantum-dot/organic semiconductor composites for
  radiation detection,'' \href{http://dx.doi.org/10.1002/adma.200501434}{{\em
  Advanced Materials} {\bfseries 18} no.~1, (2006) 77--79}.
  \url{http://dx.doi.org/10.1002/adma.200501434}.

\bibitem{Wang2010186}
C.~Wang, L.~Gou, J.~Zaleski, and D.~Friesel, ``Zns quantum dot based
  nanocomposite scintillators for thermal neutron detection,''
  \href{http://dx.doi.org/10.1016/j.nima.2010.07.032}{{\em Nuclear Instruments
  and Methods in Physics Research Section A: Accelerators, Spectrometers,
  Detectors and Associated Equipment} {\bfseries 622} no.~1, (2010) 186 --
  190}.
  \url{http://www.sciencedirect.com/science/article/pii/S0168900210016049}.

\bibitem{qdot}
L.~Winslow and R.~Simpson, ``Characterizing quantum-dot-doped liquid
  scintillator for applications to neutrino detectors,'' {\em Journal of
  Instrumentation} {\bfseries 7} no.~7, (2012) P07010.
  \url{http://stacks.iop.org/1748-0221/7/i=07/a=P07010}.

\bibitem{qdot_trans}
D.~L. Klein, R.~Roth, A.~K.~L. Lim, A.~P. Alivisatos, and P.~L. McEuen, ``A
  single-electron transistor made from a cadmium selenide nanocrystal,'' {\em
  Nature} {\bfseries 389} no.~6652, (10, 1997) 699--701.
  \url{http://dx.doi.org/10.1038/39535}.

\bibitem{qdot_photodetect}
F.~Prins, M.~Buscema, J.~S. Seldenthuis, S.~Etaki, G.~Buchs, M.~Barkelid,
  V.~Zwiller, Y.~Gao, A.~J. Houtepen, L.~D.~A. Siebbeles, and H.~S.~J. van~der
  Zant, ``Fast and efficient photodetection in nanoscale quantum-dot
  junctions,'' \href{http://dx.doi.org/10.1021/nl303008y}{{\em Nano Letters}
  {\bfseries 12} no.~11, (2012) 5740--5743}.
  \url{http://pubs.acs.org/doi/abs/10.1021/nl303008y}.

\end{thebibliography}\endgroup
\end{document}